\documentstyle[11pt]{article}
\renewcommand{\baselinestretch}{1.3}

\begin{document}

\begin{center}

{\Large {\bf Large-space shell-model calculations for light nuclei}}

\vspace{0.3in}

D. C. Zheng$^1$, J. P. Vary$^2$, and B. R. Barrett$^1$\\

\end{center}
\begin{small}

\noindent
$^1${\it Department of Physics, University of Arizona,
	Tucson, Arizona 85721}

\noindent
$^2${\it Department of Physics and Astronomy, Iowa State University,
	Ames, Iowa 50011}
\end{small}

\thispagestyle{empty}

\vspace{0.2in}

\begin{abstract}
An effective two-body interaction is constructed from a new
Reid-like $NN$ potential for a large no-core space consisting
of six major shells and is used to generate the shell-model
properties for light nuclei from $A$=2 to 6.
For practical reasons, the model space is partially truncated for
$A$=6. Binding energies and other physical observables
are calculated and they compare favorably with experiment.
\end{abstract}

\vspace*{0.2in}

{\small Suggested heading: Nuclear Structure}

\pagebreak

\section{Introduction}
In traditional nuclear shell-model calculations, only a few particles
or holes with respect to a closed shell are treated as active within a
restricted model space. In a well-studied example,
$^{18}\mbox{O}$, the model space contains one major shell,
the $1s$-$0d$ shell, with two valence nucleons.
These calculations require
effective-interaction matrix elements along with calculated or empirical
single-particle (s.p.) energies as input.
The effective interaction could either be
``phenomenological'' (see, for example, Refs.\cite{ck,wilsd,brown})
or ``realistic''(see, for example, Refs.\cite{kb18,kb40}),
depending on how it is obtained.
Both types of effective interactions have been substantially used with
success, when good agreement with experimental spectra
is taken as the criterion.
A phenomenological interaction might be obtained by fitting selected
energy spectra and electromagnetic properties of the nuclei in the region of
interest.

In the case of a realistic interaction, which is
our main concern in this work, one usually starts with the
Brueckner reaction matrix $G$ \cite{brueckner} (i.e., ladder diagrams)
calculated from a realistic nucleon-nucleon ($NN$) potential for a
specified model space,
and evaluates other diagrams (e.g. folded diagrams and/or
core-polarization diagrams \cite{eo}) using the $G$ matrix.
This renormalization procedure is incomplete, however, because
so far, the core-polarization diagrams can only be included to,
at most, the third order in the perturbation-theory expansion
\cite{bk,hom,sandel}.
The incompleteness here presents a serious problem
because convergence has not been attained within the lowest few orders
of the perturbation expansion \cite{bk,es}.
Similar uncertainties exist when calculating
the effective operators \cite{eo} to be used in the model space.

Recently, attempts \cite{vanH1,vanH2,pb,muether,lighta,veff}
have been made to avoid the above difficulty by adopting a no-core
model space in which all nucleons in a nucleus are treated as active.
It is considerably simpler to obtain the effective interaction
for a no-core space since there are no hole lines and
the complicated core-polarization processes are absent.
Consequently, one is left
with the ladder diagrams and the folded diagrams for the effective
interaction which may include effective many-body contributions.

Within the concept of no-core calculations, it is important to distinguish
two cases. In the case we call a ``full'' no-core calculation,
one selects a set of $d$ model-space s.p. states
and then generates the configurations where all nucleons can occupy all
orbitals in all possible, Pauli-allowed, ways. In an ``$N_{\rm max}\hbar\Omega$
truncated'' no-core calculation, only those configurations are
retained from the full no-core case in which there are
up-to and including $N_{\rm max}\hbar\Omega$ excitations of the lowest
unperturbed configuration (in harmonic oscillator notation) of the
$A$ nucleons.

To be more specific, let us consider a $2\hbar\Omega$ (i.e., $N_{\rm max}$=2)
shell-model diagonalization for $^{6}\mbox{Li}$
in a no-core model space consisting of the lowest four major shells
($0s$, $0p$, $1s$-$0d$ and $1p$-$0f$). In this case,
the configuration with a hole in the $0s_{\frac{1}{2}}$ shell and a particle
in the $1s$-$0d$ shell [i.e. $(0s)^3(0p)^2(1s0d)^1$] is allowed.
The configuration [i.e. $(0s)^2(0p)^4$] is also taken into account.
However, one cannot claim to have performed
a ``full'' no-core calculation because
only one or two, out of four $0s$ nucleons, are allowed to be excited
to the higher shells. Namely, in this $2\hbar\Omega$ truncated calculation,
not all nucleons are active, and there still is a partially inert core.

On the other hand, if one includes $2s$-$1d$-$0g$ and $2p$-$1f$-$0h$
major shells and performs a $4\hbar\Omega$ calculation for
the same nucleus, the configuration $(0p)^6$ will be allowed,
leaving no nucleons in the ``core'' orbital $0s_{\frac{1}{2}}$.
Although such a calculation is still restricted,
it is surely more complete than the $2\hbar\Omega$ calculation.
It is not currently possible to actually carry out a full no-core
calculation in many cases we want to investigate.
Our hope is that as $N_{\rm max}$ increases, the results will converge and
approach those of the full no-core calculation.

Another practical issue of working with an $N_{\rm max}\hbar\Omega$
truncated no-core calculation is that this facilitates the
accurate treatment of the spurious center-of-mass (c.m.)
motion. If $(N_0+1)\hbar\Omega$ is defined as the minimum
s.p. excitation energy needed to lift a nucleon
to the lowest unperturbed state outside the model space
($N_0$=4 for the $^6\mbox{Li}$ example above in the model space
through the $2p$-$1f$-$0h$ shell), then, for
$N_{\rm max}=N_0$, it is possible to obtain no-core shell-model
wavefunctions free of spurious c.m. motion.

It is an ultimate goal of the nuclear shell model to be able to start
with a realistic $NN$ potential and
obtain unambiguous and converged results against the
changes in the size of the model space and in
the choice of the unperturbed basis. Convergence with model-space size
means convergence with increasing $N_{\rm max}$
and increasing $d$ --- a dual convergence criterion.

Encouraging results have been obtained
recently for very light nuclei in Ref.\cite{ceul}, by Ceuleneer {\it et al.}
They have performed a
shell-model calculation for the $T$=0 states in $^4\mbox{He}$, where
up to $10\hbar\Omega$ excitations from the basic configuration
[$(0s_{\frac{1}{2}})^4$] are allowed. The only input to the calculation
is a set of two-body matrix elements ({\sc tbme}) of a modified
Sussex interaction.
Since this effective interaction does not have a theoretically derived
model-space dependence, they multiplied all two-body matrix elements
by a model-space dependent parameter which is adjusted to
get the correct binding energy.
The step of deriving the dependence of this parameter on the model
space size is now required to complete the dual convergence test.

In this work, we will adopt a large no-core harmonic-oscillator (HO)
model space consisting of 6 major shells (from $0s$ to $2p$-$1f$-$0h$).
We will consider several light nuclei ranging from $A$=2 to $A$=6.
An effective interaction will be constructed for the above model space
from a new Reid-like $NN$ potential (Reid93) provided by the
Nijmegen group \cite{nijm}. Note that we will use
effective interactions constructed in exactly the same manner
for all the nuclei considered here.
We will follow an approach that favors the more accurate treatment
of the spurious c.m. motion and attempts to minimize the neglect
to the two-body ``ladder'' scattering procedures. We have
designed an even more
accurate approach along these lines which
involves excitation-dependent effective interactions and will be
reported in a future work \cite{haxton}.

It is an established fact that for a small model space, a mass-dependent
two-body effective interaction gives an overall
better description \cite{wilsd,bw}.
But we anticipate that such a mass dependence will become weaker as
the size of the no-core model space is increased.
Similarly, we expect the effective many-body
forces to decrease with a increasing model space.
Indeed, if an infinitely large model space is
used, the effective interaction reverts back
to the $NN$ potential $v$, whose matrix elements are clearly independent
of the nucleus under consideration.
Throughout the remainder of this work, we assume the model space
is sufficiently large and the s.p. basis is sufficiently
realistic that we can neglect the effective many-body interactions.
This will be investigated in a future effort, which also addresses the rate
our methods approach the goal of satisfying the dual convergence criteria.

\section{Effective Interaction} \label{effi}
For a no-core model space, the core-polarization diagrams are not present,
and the {\it two-body} effective interaction is simply the
$G$ matrix \cite{brueckner} plus the folded diagrams series \cite{kuof}.
The $G$ matrix is the sum of the ladder diagram series
which represents the multiple scattering
processes of two nucleons in a nuclear medium.
We continue to follow our philosophy given in Ref.\cite{veff} for the no-core
$G$-matrix in large spaces which treats two-particle scattering
via a realistic $NN$ interaction $v_{12}$ in an ``external'' field,
$u$, which is provided by the other nucleons in the same nucleus. Thus,
we write
\begin{eqnarray}
G(\omega) &=& v_{12} + v_{12}
		\frac{Q}{\omega-(h_1+h_2)} G(\omega) \nonumber \\
          &=& v_{12} + v_{12} \frac{Q}{\omega-(h_1+h_2+v_{12})} v_{12},
					\label{G}
\end{eqnarray}
where $h=t+u$ is the one-body Hamiltonian and $u$ is the nuclear
mean field. The quantity $\omega$ is the starting energy, which
represents the initial energy of the two in-medium nucleons.
The Pauli operator $Q$ excludes the scattering of the two nucleons into
the intermediate states which are inside the model space.
It is therefore related to the choice of the model space and will be specified
in the next section.

A rigid prescription for the nuclear mean field $u$ is not necessary since
the results will be independent of $u$ once the dual convergence criteria
are satisfied. In most practical calculations, $u$ is approximated by
a one-body potential of a simple and convenient form.
The two most common choices for $u$ are a shifted HO potential
and zero:
\begin{eqnarray}
u(r) &=& -V_0 + \frac{1}{2}m\Omega^2r^2, \label{HO} \\
u(r) &=& 0.				\label{pw}
\end{eqnarray}
The latter choice corresponds to a plane-wave basis. Some hybrid
approaches have been developed which use oscillator states for low-lying
orbitals and plane waves, orthogonalized to the oscillator states, for all
the remaining states \cite{sauer,ttskuo}.

Although a shifted HO potential (\ref{HO}) does not have the expected
asymptotic behavior of vanishing exponentially at large $r$,
it was argued in Ref.\cite{veff} that the shape of the assumed $u$
at large $r$ might not be very important since, except for some
weakly bound states, nucleons are unlikely to move far beyond the
nuclear mass radius.

One may further notice from Refs.\cite{kb18,bhm} that the two seemingly
very different one-body potentials in Eqs.(\ref{HO},\ref{pw})
actually led to rather similar $G$-matrix elements, provided one makes
a careful choice for the starting energy (related to the
choice of $u$). Note that the constant shift $V_0$ in Eq.(\ref{HO})
is more a matter of convenience, as a shift of $2V_0$
can be made in the starting energy $\omega$ in Eq.(\ref{G}), i.e.,
\begin{equation}
\omega = \omega'-2V_0,  	\label{shift}
\end{equation}
to cancel out $V_0$ in the energy denominator \cite{veff}.

In this work, we will approximate the nuclear mean field
by the HO potential (\ref{HO}) not only because this simplifies the $G$-matrix
calculation \cite{bhm,vy}, but, more importantly,
for the reason that this makes possible an exact removal
of the effects of the spurious c.m. motion from our many-body
wavefunctions.
Once the $G$ matrix $G(\omega)$ is obtained as a function of the starting
energy, it will not be difficult to evaluate the folded diagrams
using the techniques developed by Kuo and Krenciglowa \cite{kuo} and by
Lee and Suzuki \cite{ls} and to obtain a starting-energy-independent
effective two-body interaction (denoted by $v^{(2)}_{\rm eff}$).

One must bear in mind that $v^{(2)}_{\rm eff}$
obtained in this procedure depends on
the assumption made for the one-body potential
in the $G$-matrix calculation. Especially in cases when the model spaces
are small and we are further from satisfying the dual convergence
criteria, it is important to use a $u$ that sensibly
represents the nuclear mean
field so as to minimize the neglected many-body interactions
\cite{barrett} and higher than linear order ``--$u$'' insertions.

In Ref.\cite{veff} it is shown that $v^{(2)}_{\rm eff}$ can be well
approximated by the $G$ matrix calculated at starting energies
which are related to the initial unperturbed energy of the two
nucleons in the ladder scattering processes in a simple way:
\begin{equation}
\omega' = \omega +2V_0 = \epsilon_a+\epsilon_b + \Delta,
					\label{omega}
\end{equation}
where $\epsilon=(2n_r+l+3/2)\hbar\Omega$ are the HO s.p. energies
($a$ and $b$ are the s.p. states that the two nucleons
initially occupy). Such a state-dependent choice for $\omega'$ will
lead to a non-hermitian $G$ matrix, but the non-hermicity is found
to be small. The quantity $\Delta$ signifies the interaction
energy between the two nucleons. In a specific application to
$^6\mbox{Li}$, it has been shown \cite{veff} that
for $\hbar\Omega$=18 MeV, a value of -21 MeV for $\Delta$
results in $G(\omega')$ which is
an excellent approximation to $v^{(2)}_{\rm eff}$.

In this work, we will follow Ref.\cite{veff} and adopt
the average of $G(\omega')$ and its conjugate
calculated for a HO basis with $\hbar\Omega$=14 MeV
with $\omega'$ given by Eq.(\ref{omega})
as our approximation to $v^{(2)}_{\rm eff}$.
The parameter $\Delta$ is chosen to yield the experimental
binding energy. Initially, one might expect that
different values of $\Delta$ have to be used for different nuclei.
But, quite surprisingly, we find that good agreement with experimental
observables can be obtained with a nucleus-independent
value of $\Delta$ (--35 MeV).

Our shell-model Hamiltonian will now be written as
\begin{equation}
H_{\rm SM} = \left(\sum_{i=1}^A t_i -T_{\rm c.m.}\right)
	+ \sum_{i<j}^A G_{ij} + V_{\rm Coulomb}
	+ \lambda (H_{\rm c.m.}-\frac{3}{2}\hbar\Omega),	\label{hsm}
\end{equation}
where $t_i={\bf p}_i^2/(2m)$ are the one-body kinetic energies,
$T_{\rm c.m.}=(\sum_i {\bf p}_i)^2/(2mA)$
the c.m. kinetic energy and $V_{\rm Coulomb}$ the Coulomb
interaction. The proton and neutron masses are taken to be the same.
The last term (with $\lambda$=10) in the above equation forces
the c.m. motion of the low-lying states in the calculated spectrum
to be in its lowest HO configuration.

We remark that our calculations involve no free parameters other than
those used in calculating the $G$ matrix, $\hbar\Omega$ and $\Delta$.
Moreover, these two parameters are fixed at 14 MeV and -35 MeV, respectively,
for all nuclei considered in the present work.

We emphasize that in a no-core calculation, we are attempting to derive all
shell-model properties from an underlying Hamiltonian, $H_{\rm SM}$.
Thus, there are no phenomenological s.p.
energy terms in $H_{\rm SM}$.

\section{Results and Discussions}
As previously mentioned, we use
a no-core model space containing the lowest six HO major shells
with $\hbar\Omega$=14 MeV. For $A\le 4$, we allow all
$0\hbar\Omega$ through $7\hbar\Omega$ configurations within the model space.
For $A>4$, we allow all $0\hbar\Omega$ through $5\hbar\Omega$ configurations.
Therefore, different $Q$ operators have to be used in Eq.(\ref{G})
for $A\le 4$ and $A>4$:
\begin{small}
\begin{eqnarray}
{\rm For} \hspace{0.1in} A\le 4: \hspace{0.2in}
Q &=& 1 \hspace{0.2in} {\rm for} \hspace{0.1in} n_1\ge 6,\; n_2\ge 6,
	\hspace{0.1in} {\rm or}  \hspace{0.1in} n_1+n_2 \ge 8, \label{q1}\\
  &=& 0 \hspace{0.2in} {\rm otherwise};  \nonumber \\
{\rm For} \hspace{0.1in} A>4: \hspace{0.2in}
Q &=& 1 \hspace{0.2in} {\rm for} \hspace{0.1in} n_1+n_2 \ge 6, \label{q2}\\
  &=& 0 \hspace{0.2in} {\rm otherwise}. \nonumber
\end{eqnarray}
\end{small}

\noindent
In the above equations, $n$=$2n_r$+$l$ is the principal quantum number for
the HO s.p. states. It starts from 0 with
$n$=0 representing the first major shell ($0s$).
For $A$=6, due to the computer memory limitation, the $n$=5 shell
contains only the $p$ orbitals; the $f$ and $h$ orbitals are left
outside the model space.

The shell-model matrix diagonalizations are performed for the
Hamiltonian $H_{\rm SM}$ in Eq.(\ref{hsm}) using the
Many-Fermion-Dynamics code \cite{mfd}. The results are given
in Table I, which we will discuss in the following subsections.
The experimental results given in Table I are taken from
Ref.\cite{exp3} for $A$=3, Ref.\cite{exp4} for $A$=4 and
Ref.\cite{expA} for $A$=5 and 6.

\subsection{Binding Energies}
It is possible to obtain exact or nearly exact results for ground-state
energies of the lightest nuclei by solving the Schr\"odinger
or Faddeev \cite{faddeev} equations for realistic $NN$ interactions.
This has been done for
the ground states of $^3\mbox{H}$, $^3\mbox{He}$ and $^4\mbox{He}$
\cite{chen,carlson,kg,wiringa1}.
Even for $^5\mbox{He}$ and $^6\mbox{Li}$, preliminary results obtained
by Wiringa in variational Monte Carlo calculations
have appeared \cite{wiringa2}.
Unlike the few-body approaches in which one obtains almost exact
results, at least for the ground state,
the effective-interaction shell-model approach involves some
uncertainties due to the truncation of the space and the approximation made
in calculating the effective interaction for the truncated space.
Consequently, the shell-model approach to the above light nuclei, although
being able to calculate for a given set of quantum numbers
the excited states as easily as the lowest state,
cannot match the few-body approach in the accuracy of the results
for the ground state. Nevertheless, our ultimate goal is to
satisfy the dual covergence criteria. By using a large no-core
model space along with a reasonable effective interaction,
we hope to demonstrate that, in spite of its present limitations,
the effective Hamiltonian
approach gives a useful description of the low-lying states in light nuclei.

Our results are encouraging as can be seen from Table I.
The calculated binding energies for the deuteron, triton, $^4\mbox{He}$,
$^5\mbox{He}$ and $^6\mbox{Li}$ are 2.103, 8.589, 28.757, 25.960 and
30.648 MeV, respectively,
agreeing quite well with the corresponding experimental values
of 2.225, 8.482, 28.296, 27.410 and 31.996 MeV. Of course, it is more
relevant to compare our results to those obtained in the exact
few-body approaches using the same potential. These approaches show
that existing realistic $NN$ potentials underbind light nuclei with $A>2$.
Our calculations involve a free parameter $\Delta$ which
has been fixed at -35 MeV for all the nuclei considered.
For $A>2$, we can obtain smaller binding energies (in better agreement
with exact calculations) by decreasing $\Delta$ (i.e., making it more negative)
since the binding energies decrease monotonically with the
decreasing $\Delta$.
Our adoption of a $\Delta$ value that fits experimental binding energies
stems from an assumption that our neglected effective many-body
forces and other corrections can largely cancel the neglected (and largely
unknown) true many-body forces.

It is worth mentioning that for the two-body system, the deuteron,
it is possible to obtain exact results even with the effective-interaction
approach \cite{deuteron}. Our present results for the deuteron
are not exact due to our neglect of the processes which are higher order
in $u$. Our effective interaction nevertheless gives a
reasonable deuteron binding energy. In section \ref{moments},
we will further show that the deuteron magnetic dipole and electric
quadrupole moments also come out well.

\subsection{Excitation Spectra}
For the deuteron and the triton, we obtain only one bound state in the
calculations, agreeing with experiment and with exact calculations.
For the deuteron, the lowest state in the continuum is a
$J^{\pi}$=$0^+$, $T$=1 state, which is unbound by 1.65 MeV
(i.e. 3.75 MeV above the ground state).
For the triton, the lowest excited state is a
$J^{\pi}$=${\frac{5}{2}}^-$, $T$=$\frac{1}{2}$ state, unbound by 4.13 MeV.
It has a nearly degenerate $J^{\pi}$=${\frac{1}{2}}^-$,
$T$=$\frac{1}{2}$ state,
unbound by 4.28 MeV. The $T$=$\frac{3}{2}$ states are even higher in energy.
Therefore, these results do not support $nn$, $pp$,
$nnn$ or $ppp$ bound states.

For $^4\mbox{He}$, the experimental level sequence of the low-lying
negative-parity states is correctly reproduced. The excitation energies
are consistently higher than the experimental results \cite{exp4} by about
2 to 3 MeV. These results are clearly an improvement over
those obtained in our previous study \cite{lighta}. In that study,
the excitation energies of these same states were
obtained in a smaller model space,
including only four major shells, and were found to be
as much as 6 MeV too high
when compared with experiment (see Table I in Ref.\cite{lighta}).
The better results we obtain here should be attributed mainly to
the larger model space and the improved $NN$ interaction.
{}From a theoretical viewpoint, we have also improved the
$G$ matrix by using a state-dependent starting energy
of Eq.(\ref{omega}) (rather than at a constant starting energy as in
Ref.\cite{lighta}) which better approximates the full effective
interaction \cite{veff}.

We obtain the first excited state ($J^{\pi}$=$0^+$, $T$=0) in $^4\mbox{He}$
at an excitation energy of 26.135 MeV. This is about 6 MeV higher than
experiment but it is about 7.7 MeV lower than the previous
result (33.807 MeV) for the four-major-shell space \cite{lighta}.
Again, the larger model space used in this work is
largely responsible for the decrease in energy.
A more accurate description of this state will require an even larger space.
Indeed, in Ref.\cite{ceul} where a modified Sussex interaction is used,
excellent agreement with experiment is obtained for this state only
when up to $10\hbar\Omega$ configurations are included.

The calculated ground state in $^4\mbox{He}$ is dominated by the
$(0s)^4$ configuration but it has a considerable amount of
``$1p$-$1h$'' configuration $(0s)^3(1s)^1$ and ``$2p$-$2h$'' configuration
$(0s)^2(0p)^2$. The $0^+_2$ state is dominated by the $(0s)^3(1s)^1$
configuration while the $(0s)^2(0p)^2$ and $(0s)^4$ components are also
quite significant.

The occupancies of
the various model-space orbitals in the $0^+_1$ and $0^+_2$ states are
\begin{small}
\begin{eqnarray}
|^4{\rm He}: 0^+_1\rangle &=& (0s)^{3.525} (0p)^{0.160} (sd)^{0.200}
	({\rm other\; shells})^{0.114}; \\
|^4{\rm He}: 0^+_2\rangle &=& (0s)^{2.679} (0p)^{0.521} (sd)^{0.716}
	({\rm other\; shells})^{0.084}.
\end{eqnarray}
\end{small}
Relative to the ground state, the $0^+_2$ state has only about
50\% of the ``breathing mode'' $(0s)^{-1}(1s)^1$.
However, this result depends on the choice of the
s.p. basis. If, for example, a Hartree-Fock basis were used,
the oscillator-basis $s$-states would mix to produce the HF $s$-states
(e.g. $0s_{\rm HF}$ and $1s_{\rm HF}$ {\it etc.}) so that the admixture of
the $(0s_{\rm HF})^3(1s_{\rm HF})^1$ component in the ground state would
likely be much smaller. This would lead to a larger amount of
$(0s_{\rm HF})^{-1}(1s_{\rm HF})^1$ in the $0^+_2$ state.

Note that although our model space is not sufficiently large to reproduce
the first excited $0^+$ state in $^4\mbox{He}$ at the experimental
energy, it does a fairly good job for the ``$1\hbar\Omega$'' states.
This gives us confidence in the results for the low-lying states
in $^5\mbox{He}$, which we present below.

The first excited state in $^5\mbox{He}$
($J^{\pi}$=${\frac{1}{2}}^-$, $T$=$\frac{1}{2}$)
is obtained at an energy of 3.112 MeV, within the range of 3 to 5 MeV
given in Ref.\cite{expA}. The low-lying positive-parity states are also of
interest. Experimentally, there is a famous
$J^{\pi}$=${\frac{3}{2}}^+$ state at 16.75 MeV.
It has a dominant $(0s)^3(0p)^2$ configuration and can be thought
of as the ground state of $^6\mbox{Li}$ coupled to a $0s$ hole.
This state corresponds to our calculated 20.445 MeV state in Table I,
which has the following occupation probabilities:
\begin{equation}
|20.445{\rm MeV}: {\frac{3}{2}}^+_2\rangle =
	(0s)^{2.766} (0p)^{1.906} (sd)^{0.237} ({\rm other\; shells})^{0.091},
\end{equation}
to be compared to the occupation probabilities for the ground state of
$^6\mbox{Li}$
\begin{equation}
|^6\mbox{Li}: 1^+_1\rangle =
	(0s)^{3.631} (0p)^{2.061} (sd)^{0.222} ({\rm other\; shells})^{0.087}.
\end{equation}
Note that the fact that the calculated energy of this
$J^{\pi}$=${\frac{3}{2}}^+$ state is
about 3.7 MeV higher than experiment is
more or less consistent with what we have seen in
the case of $^4\mbox{He}$ where the ``$1\hbar\Omega$'' low-lying
negative-parity states came out about 2 to 3 MeV higher than experiment.
It will be interesting to track the energies of these states
as well as the excited states in $^4\mbox{He}$
with increasing model-space size.

Above the two s.p. states (the ground ${\frac{3}{2}}_1^-$
state and the first excited ${\frac{1}{2}}^-_1$ state)
and below the 20.445 MeV ${\frac{3}{2}}_2^+$ state, our calculation also
gives three positive-parity ``$1\hbar\Omega$'' states,
a ${\frac{1}{2}}^+$ state at 7.437 MeV and nearly degenerate
${\frac{5}{2}}^+$ and ${\frac{3}{2}}^+$ states at 14.206 and 14.439 MeV,
respectively. These states are dominated
by the configurations $(0s)^4(sd)^1$ and $(0s)^3(0p)^2$.
The occupancies of the orbitals in the model space are:
\begin{small}
\begin{eqnarray}
|7.437{\rm MeV}: {\frac{1}{2}}^+_1\rangle &=&
     (0s)^{3.219} (0p)^{0.933} (sd)^{0.681} ({\rm other\; shells})^{0.167}; \\
|14.206{\rm MeV}: {\frac{5}{2}}^+_1\rangle &=&
     (0s)^{3.197} (0p)^{0.928} (sd)^{0.671} ({\rm other\; shells})^{0.204}; \\
|14.439{\rm MeV}: {\frac{3}{2}}^+_1\rangle &=&
     (0s)^{3.153} (0p)^{1.011} (sd)^{0.623} ({\rm other\; shells})^{0.213}.
\end{eqnarray}
\end{small}

\noindent
There have been previous
theoretical predictions \cite{expA,millener} that there is
a ${\frac{1}{2}}^+$ state at about 5 MeV and
${\frac{3}{2}}^+$ and ${\frac{5}{2}}^+$ states
at about 12 MeV. These predictions have not been fully
confirmed experimentally, but they are well supported by our results,
again, keeping in mind that our calculated ``$1\hbar\Omega$'' states are
probably about two or three MeV too high.

In addition to the above low-lying states, we have also listed in Table I
a few other bound states of $^5\mbox{He}$
which have an energy not much higher than the experimental 16.75 MeV state.

The low-lying energy spectrum of $^6\mbox{Li}$ obtained in this calculation
does not show much improvement over that in Ref.\cite{lighta}. It again
appears to be more spread-out than the experimental spectrum.

\subsection{M1 and E2 Moments}   \label{moments}
Since we are using a large no-core model space, we choose to use
bare operators ($e_p$=1, $e_n$=0, $g_p^s$=5.586, $g_n^s$=-3.826,
$g_p^l$=1.0, $g_n^l$=0.0) to calculate the magnetic dipole (M1) and
electric quadrupole (E2) moments in leading approximation.
The calculated results are also given
in Table I. It should be emphasized that only the
nucleonic degrees of freedom are taken into account
in calculating these moments.
Proper considerations have to be given to the effects
of the meson exchange currents (MEC) before critical conclusions
can be drawn from the comparison of
the calculated moments (especially the M1 moment) with data.

The calculated M1 moment $\mu$ for the deuteron is $0.857\mu_N$.
This agrees with the experimental result of $0.8574\mu_N$. However,
this fortuitous agreement will be vitiated to the extent
that the ignored MEC contribution is significant.
Even if the MEC effect is
negligible, the value that we obtained for the deuteron M1 moment
is not theoretically exact. This is made evident in the discussion below.

The deuteron M1 moment is related to the $D$-state
probability $P_D$ as:
\begin{equation}
\mu(^2{\rm H}) = P_S \mu(^3S_1) + P_D \mu(^3D_1)
               = (1-P_D)0.880 + P_D 0.310\; (\mu_N).
\end{equation}
With this equation, a calculated value of 0.857$\mu_N$
for $\mu(^2{\rm H})$ leads to $P_D$=4.0\%. However,
the exact $P_D$ for the Reid93 potential is in fact 5.7\%
\cite{friar}, implying a $\mu(^2{\rm H})$ of 0.848$\mu_N$.
We, therefore, see that the tensor force is somehow weakened when we
go from the bare $NN$ potential to the effective shell-model interaction in
Eq.(\ref{hsm}) for our no-core model space.
This infers the size of the neglected contribution
to the magnetic moment operator arising in the theory of effective
operators.
It has been shown in Ref.\cite{zz} that the tensor force strength
can be further reduced by core-polarization diagrams (mainly the Bertsch
bubble diagram \cite{Bertsch}) that one must take into account when
calculating the effective interaction for a small, one-major-shell,
model space outside an inert core.

The calculated deuteron quadrupole moment $Q$ is 0.242$e{\rm fm}^2$,
somewhat smaller than the experimental value of 0.286$e{\rm fm}^2$.
This agrees with the above observation that the effective tensor force in our
no-core shell-model interaction is weaker than that in the original
NN potential. The reduced quadrupole moment may also arise from the fact that
its operator involves a radial dependence ($r^2$) which needs
to be renormalized when we truncate the infinite Hilbert space to
our finite-size no-core HO model space.
Thus we reason that, for our model space, the renormalization
effects are larger for the E2 operator than for the M1 operator
which does not have a radial dependence.

The need for using an effective operator to evaluate the root-mean-squared
(rms) radius (or any other observable that involves it) is evident from
Table I, where the calculated rms {\em point} radius
$\sqrt{\langle r^2_p\rangle}$ for the proton in the deuteron is
1.488 fm, significantly smaller than the experimental value of 1.95 fm.
The large renormalization of the rms radius operator required for the
deuteron is not surprising since it is a very loosely bound system, the
wave function obtained in the truncated HO model space does not represent
the exact wave function very well.
The calculated $\sqrt{\langle r^2_p\rangle}$ value for $^6\mbox{Li}$
is also smaller than the experimnetal value.
However, the results of $\sqrt{\langle r^2_p\rangle}$ for
$^3\mbox{H}$ and $^4\mbox{He}$ are in good agreement with experiment.
Note that we have evaluated these rms radii with ``intrinsic''
wavefunctions so the quoted results are free of spurious c.m.
contributions.

Our calculated M1 moment for the triton is 2.659$\mu_N$, about 11\%
smaller than the experimental value of 2.979$\mu_N$.  To a large extent,
this discrepancy may be explained
by the MEC effects that we have not taken
into account. Indeed, in Ref.\cite{wiringa2}, it is shown that the inclusion
of the MEC effects in a model-dependent way
leads to a 14\%'s increase in the triton M1 moment from 2.588$\mu_N$ to
3.010$\mu_N$, in close agreement with experiment.

For the ground state of $^5\mbox{He}$, the calculated M1 and E2 moments
are -1.864$\mu_N$ and -0.332$e{\rm fm}^2$, respectively. Again,
the MEC effects have to be considered when comparing these results
with experimental data, which, to our knowledge, are not available.

It has been difficult in the past for theory to reproduce the
E2 moment for the ground state of $^6\mbox{Li}$.
However, the calculated E2 moment is  -0.116$e{\rm fm}^2$, which is
remarkably close to the experimental value of -0.082$e{\rm fm}^2$.
Our calculated M1 moment is 0.851$\mu_N$, which is
about 3.5\% higher than the experimental result of 0.822$\mu_N$.

\subsection{Effects of the Coulomb Interaction}

Since we include the Coulomb interaction, the isospin symmetry is not strictly
conserved. But the isospin impurity
caused by the Coulomb interaction is generally very small. For the
bound states of $^3\mbox{H}$ and $^3\mbox{He}$, the calculated values for
isospin
\begin{equation}
T_{\rm calc} = \frac{\sqrt{4\langle {\hat{T}}^2\rangle+1} -1 }{2}
\end{equation}
are 0.500000 and 0.500022, respectively. Note that $^3\mbox{H}$
has only one proton so isospin is still a good quantum number.
In $^3\mbox{He}$, the calculated isospin shows only a 0.0044\% deviation
from the half-integer value.
$T_{\rm calc}$ is 0.000046 for the ground state of $^4\mbox{He}$; it is
0.500016 and 0.500024 for the ground states of $^5\mbox{He}$
and $^5\mbox{Li}$, respectively. The small isospin impurity
for the ground states in these nuclei is due to the fact
that all these states do not have any nearby state with the same $J^{\pi}$
but a different $T$. From perturbation theory, one knows
that the relatively weak Coulomb interaction will not induce much
isospin mixing to these isolated states.

The Coulomb interaction has sizable effects on the absolute
energies of the system,
as is well known. Our calculation shows that,
due to the Coulomb repulsion, the
binding energy of $^3\mbox{He}$ is 0.725 MeV less than that of
$^3\mbox{H}$ and the binding energy of $^5\mbox{Li}$ is 1.024 MeV
less than that of $^5\mbox{He}$. The experimentally observed differences
in the binding energies for the above two pairs are
0.764 and 1.073 MeV, respectively. They
are quite close to our calculated values, as one might expect
since the Coulomb interaction is a perturbation in these light systems.
Nevertheless, our results for the Coulomb energy are model-dependent
[in that the matrix elements of the Coulomb interaction
in the shell-model Hamiltonian (\ref{hsm}) were evaluated using
a HO basis and possible renormalization corrections from the excluded
space were ignored]. A smaller Coulomb effect of about 0.74 MeV
was obtained in a more model-independent analysis \cite{friar3}
for the $^3\mbox{H}$--$^3\mbox{He}$ pair. It is also believed
that other charge-symmetry breaking effects contribute
to the difference between the binding energies of
$^3\mbox{H}$ and $^3\mbox{He}$ as well \cite{coon}.

\section{Conclusions}
In this work, we have constructed an effective interaction for a
six-major-shell no-core model space from a new, Reid-like,
NN potential (Reid93) from the Nijmegen group \cite{nijm}.
The effective interaction has been applied to calculate
nuclear structure properties for a few light nuclei, ranging from
the deuteron to $^6\mbox{Li}$.
The results are very encouraging. Not only are the binding energies
of these nuclei well reproduced,
the energy spectra are also in good agreement with experiment.
In particular, the experimental level sequence of
the low-lying negative-parity states in $^4\mbox{He}$
is correctly reproduced, although the excitation energies are about 2 to 3
MeV higher than experiment.
Based on our current and previous efforts, we expect that this
discrepancy will be reduced as we more closely satisfy the dual
convergence criteria --- convergence against increasing $N_{\rm max}$ and $d$,
where $N_{\rm max}$ signifies the highest unperturbed energy of
the configurations taken into account
and $d$ represents the number of s.p. states included in the model space.

The magnetic dipole and electric quadrupole moments, calculated
using bare operators with meson-exchange-currents effects neglected,
are also in reasonable agreement with experiment.

For $^5\mbox{He}$, in addition to the two low-lying s.p.
negative-parity states ${\frac{3}{2}}^-$ and ${\frac{1}{2}}^-$,
we have obtained a low-lying ${\frac{1}{2}}^+$ state
at about 7.4 MeV and two nearly degenerated states (${\frac{5}{2}}^+$ and
${\frac{3}{2}}^+$)
at 14.2 MeV and 14.4 MeV. The latter three, dominated by
the configuration $(0s)^4(sd)^1$, are mainly s.p. states
with one $(sd)$ neutron coupled to the ground state of $^4\mbox{He}$.
The actual energies of these predominantly
``$1\hbar\Omega$'' states could be about a few MeV lower, as in the case of
$^4\mbox{He}$. The previous
theoretical predictions of a ${\frac{1}{2}}^+$ state
at about 5 MeV and ${\frac{5}{2}}^+$ and
${\frac{3}{2}}^+$ states at about 12 MeV are
therefore well supported by our results. The 16.75 MeV state, resulted from
the ground state of $^6\mbox{Li}$ with a $(0s)$ proton removed,
is reproduced at an energy of 20.445 MeV.

The Coulomb interaction, which is
included in the calculations, accounts for the bulk part
of the differences in the experimental binding energies of mirror pairs
($^3\mbox{H}$-$^3\mbox{He}$ and $^5\mbox{He}$-$^5\mbox{Li}$).
We have also seen that the Coulomb interaction
induces a very small amount of isospin impurity to the ground states
of the light nuclei considered.

An extension of the current approach to heavier $0p$-shell nuclei will be
straightforward. Our results for $A$=2 to 6 have given us optimism
that our approach would be able to give
a good description of neighboring nuclei as well.
This is presently being investigated.

Of course, since the size of the shell-model matrix increases
quite dramatically with the increasing number of nucleons, it is unlikely
at the present time
that one can apply the no-core approach to a much heavier nucleus, like
$^{40}\mbox{Ca}$.
In this regard, the Monte Carlo shell-model approach \cite{koonin},
in which the size of the calculations increases only moderately
with the number of active nucleons, offers some promise.

\section*{Acknowledgment}
We are grateful to S.A. Coon for many useful communications and
to D.W.L. Sprung for reading the manuscript.
Two of us (B.R.B. and D.C.Z.) acknowledge
partial support of this work by the National Science Foundation,
Grant No. PHY91-03011. One of us
(J.P.V.) acknowledges partial support by the U.S.
Department of Energy under Grant No. DE-FG02-87ER-40371, Division
of High Energy and Nuclear Physics.

\begin{small}

\end{small}

\pagebreak

\renewcommand{\baselinestretch}{1.2}

\begin{small}


\noindent
{\bf Table I}. The results for $^2\mbox{H}$, $^3\mbox{H}$,
$^4\mbox{He}$, $^5\mbox{He}$ and $^6\mbox{Li}$ obtained in
large no-core (consisting of 6 HO major shells) shell-model
calculations. The experimental data are taken from Refs.\cite{exp3,exp4,expA}.
In the Table, $E_{B}$ is the binding energy (in MeV);
$E_x(J_n^{\pi},T)$ the excitation energy (in MeV) of
the $J^{\pi}_n,T$ state. The ground-state
rms {\em point} radius for protons $\sqrt{\langle r^2_p\rangle}$ (in fm),
electric quadrupole moment $Q$ (in $e {\rm fm}^2$)
and magnetic dipole moment $\mu$ (in $\mu_N$) are also listed.
The ``experimental'' $\sqrt{\langle r^2_p\rangle}$
was deduced from the charge rms radius
$\sqrt{\langle r^2_c\rangle}$ through (ignoring the neutron
charge distribution and other higher-order effects and
assuming a proton rms charge radius of 0.81 fm)
$\langle r^2_p\rangle = \langle r^2_c\rangle - 0.81^2$.

\pagebreak
\vspace*{-0.5in}

\begin{center}
\begin{tabular}{ccc|ccc} \hline\hline
Observable               &  Calc. & Exp't &
Observable               &  Calc. & Exp't \\ \hline
\multicolumn{3}{c}{\bf Deuteron} & \multicolumn{3}{|c}{\bf Triton} \\
$E_B$               &  2.103 & 2.2246   & $E_B$             & 8.589 & 8.4819 \\
$\sqrt{\langle r^2_p\rangle}$
                    &  1.653 & 1.95  &
             $\sqrt{\langle r^2_p\rangle}$             &  1.573 & 1.41--1.62\\
$\mu$               &  0.857 & 0.8573 & $\mu$              &  2.659 & 2.9790\\
$Q$                 &  0.242 & 0.2859 &
        $E_x({\frac{5}{2}}^-_1,\frac{1}{2})$               & 12.716 & unbound\\
$E_x(0^+_1,1)$      &  3.754 & unbound&
        $E_x({\frac{1}{2}}^-_1,\frac{1}{2})$               & 12.868 & unbound\\
								      \hline
\multicolumn{3}{c}{\bf $^4\mbox{He}$}
		& \multicolumn{3}{|c}{\bf $^5\mbox{He}$}\\
$E_B$               & 28.757 & 28.296 & $E_B$              & 25.960 & 27.410\\
$\sqrt{\langle r^2_p\rangle}$
                    &  1.488 & 1.46  &
  $\sqrt{\langle r^2_p\rangle}$                            &  1.659 &       \\
$E_x(0^+_1,0)$      &  0.000 & 0.00   & $\mu$              & -1.864 &        \\
$E_x(0^+_2,0)$      & 26.135 & 20.21  & $Q$                & -0.332 &        \\
$E_x(0^-_1,0)$      & 22.848 & 21.01  &
        $E_x({\frac{3}{2}}^-_1,\frac{1}{2})$               &  0.000 & 0.00   \\
$E_x(2^-_1,0)$      & 24.351 & 21.84  &
        $E_x({\frac{1}{2}}^-_1,\frac{1}{2})$               &  3.112 &$4\pm 1$\\
$E_x(2^-_1,1)$      & 25.739 & 23.33  &
        $E_x({\frac{1}{2}}^+_1,\frac{1}{2})$               &7.437&See $^{a)}$\\
$E_x(1^-_1,1)$      & 26.338 & 23.64  &
        $E_x({\frac{5}{2}}^+_1,\frac{1}{2})$              &14.206&See $^{a)}$\\
$E_x(1^-_1,0)$      & 27.337 & 24.25  &
        $E_x({\frac{3}{2}}^+_1,\frac{1}{2})$              &14.439&See $^{a)}$\\
$E_x(0^-_1,1)$      & 27.418 & 25.28  &
        $E_x({\frac{3}{2}}^+_2,\frac{1}{2})$             &20.445&16.75$^{b)}$\\
$E_x(1^-_2,1)$      & 27.905 & 25.95  &
        $E_x({\frac{3}{2}}^-_2,\frac{1}{2})$               & 21.499 & N/A \\
						\cline{1-3}
\multicolumn{3}{c|}{\bf $^6\mbox{Li}$} &
        $E_x({\frac{1}{2}}^+_2,\frac{1}{2})$               & 23.563 & N/A \\
$E_B$               & 30.648 & 31.996 &
        $E_x({\frac{7}{2}}^+_1,\frac{1}{2})$               & 23.592 & N/A \\
$\sqrt{\langle r^2_p\rangle}$ & 2.050 & 2.38 &
        $E_x({\frac{1}{2}}^-_2,\frac{1}{2})$               & 24.045 & N/A \\
$\mu$               &  0.851 & 0.822 &
        $E_x({\frac{3}{2}}^+_3,\frac{1}{2})$               & 24.398 & N/A \\
$Q$                 & -0.116 & -0.082 &
        $E_x({\frac{1}{2}}^+_1,\frac{3}{2})$               & 25.861 & N/A \\
$E_x(1^+_1,0)$      &  0.000 & 0.000  &
        $E_x({\frac{1}{2}}^+_3,\frac{1}{2})$               & 26.240 & N/A \\
$E_x(3^+_1,0)$      &  2.959 & 2.186  &
        $E_x({\frac{3}{2}}^+_4,\frac{1}{2})$               & 27.359 & N/A \\
$E_x(0^+_1,1)$      &  3.607 & 3.563  &
        $E_x({\frac{7}{2}}^-_1,\frac{1}{2})$               & 27.681 & N/A \\
$E_x(2^+_1,0)$      &  5.485 & 4.31   \\
$E_x(2^+_1,1)$      &  6.505 & 5.366   \\
$E_x(1^+_2,0)$      &  7.828 & 5.65   \\ \hline\hline
\end{tabular}
\end{center}
\noindent
\hspace*{-0.2in}
$^{a)}$ Low-lying positive-parity states (e.g. a $J^{\pi}$=${\frac{1}{2}}^+$,
$T$=$\frac{1}{2}$
state at $\sim$5 MeV and $J^{\pi}$=${\frac{3}{2}}^+$, $T$=$\frac{1}{2}$ and
$J^{\pi}$=${\frac{5}{2}}^+$, $T$=$\frac{1}{2}$ states at $\sim$12 MeV)
are predicted to exist. See Ref.\cite{expA} for more details.

\noindent
\hspace*{-0.2in}
$^{b)}$ We identify the calculated 20.445 MeV state as the experimental
16.75 MeV state, because the calculated state is dominated by
the $(0s)^3(0p)^2$ configuration.

\end{small}
\end{document}